\documentclass[useAMS,usenatbib]{mn2e}
\usepackage{graphicx,natbib,amssymb}
\newcommand{\msun}{\mbox{\rm M$_{\odot}$}}

\title[Optical spectroscopy of high proper motion stars]{Optical spectroscopy of high proper motion stars: new M dwarfs within 10 pc and the closest pair of subdwarfs}
\author[C. Reyl\'e et al.]{C. Reyl\'e$^{1}$\thanks{E-mail:
celine@obs-besancon.fr}, R.-D. Scholz$^{2}$, M. Schultheis$^{1}$, A. C. Robin$^{1}$ and M. Irwin$^{3}$\\
$^{1}$CNRS UMR6091, Observatoire de Besan\c{c}on, BP 1615, 25010 Besan\c{c}on Cedex, France\\
$^{2}$Astrophysikalisches Institut Potsdam, An der Sternwarte 16, Potsdam, D-14482, Germany\\
$^{3}$Institute of Astronomy, University of Cambridge, Madingley Road, Cambridge, CB3 0HA, England}
\begin{document}

\date{Accepted . Received ; in original form }

\pagerange{\pageref{firstpage}--\pageref{lastpage}} \pubyear{}

\maketitle

\label{firstpage}

\begin{abstract}
We present spectra of 59 nearby stars candidates, M dwarfs and white dwarfs, 
previously identified using high proper motion catalogues and the DENIS database. We
review the existing spectral classification schemes and spectroscopic parallax
calibrations in the near-infrared $J$-band and derive spectral types and distances
of the nearby candidates. 42 stars have spectroscopic distances smaller than 25~pc, 
three of them being white dwarfs. 
Two targets lie within 10~pc, one M8 star at 10.0~pc (APMPM J0103-3738), and one M4 star at
8.3~pc (LP 225-57). One star, LHS 73, is found to be among the few subdwarfs lying within 20 pc.
Furthermore, together with LHS 72, it probably belongs to the closest pair of subdwarfs we know.
\end{abstract}


\section{Introduction}

Recent discoveries of cool objects, such as M stars or brown dwarfs, 
closer than 5 parsecs show that even the immediate solar neighbourhood 
sample is still incomplete
\citep{delfosse2001,scholz2003,teegarden2003,hambly2004}. 

\citet{henry1997} estimated that about 130 systems over 359 (36\%) are missing within 10 pc.
The missing fraction is even larger within 25 pc (63\%) with a deficit of about 3500 systems over 
the 5500 expected ones \citep{henry2002}. Statistical comparisons from the local sample
in the northern hemisphere led to a less pessimistic result, the current 10 pc sample being 
$\sim$75\% complete \citep{reid2003a}.

New surveys in the near infrared such as DENIS \citep{epchtein1997} and 2MASS 
\citep{cutri2003} provide unprecedented data 
for a systematic search for low luminosity cool dwarfs. The use of these data 
together with high proper motion catalogues is a powerful tool for discovering 
our neighbours. Hundreds of stars closer than 25 parsecs have been discovered 
this way \citep[e.g.][]{phanbao2001,phanbao2003,reid2002a,reid2002b,reyle2002,reid2003b,reid2004,hambly2004,reyle2004,lodieu2005,scholz2005}.

As spectroscopy provides much more information than photometry alone, spectroscopic observations were also carried out to identify and classify nearby stars. Recent studies dealing with sample sizes larger 
than 10 objects are cited in Table~1. All together, they revealed 490 stars in the 25 pc sample and 25 within 10 pc. However, we note that the samples investigated by the different authors
do have many stars in common so that the total number of newly
discovered neighbours is much smaller.

\begin{table}
\label{tab1}
\caption{Nearby stars identified by spectroscopy }
\begin{tabular}{lccc}
\hline
Reference 	&Sample &\multicolumn{2}{c}{Number of stars within} \\
		 	& size                        &25 pc	&10 pc \\
\hline
\citet{cruz2002}	  	&70			&28		& \\
\citet{henry2002} 	&34			&6		&2 \\
\citet{reid2003b}  	&357		&127	&9 \\
\citet{lodieu2005}	&71			&25		&3 \\
\citet{crifo2005}		&39			&31		&1 \\
\citet{scholz2005}	&322		&226	&8 \\
\citet{phanbao2006}	&45			&5		& \\
This study			&59			&42		&2 \\
\hline\\
\end{tabular}
\end{table}

In previous papers \citep{reyle2002,reyle2004}, we have determined the photometric 
distances of high proper motion stars that we cross-identified with the DENIS survey.
We have reported the discovery of 
115 nearby candidates, probably lying within 25 pc, the limit of the Catalogue of
Nearby Stars \citep[CNS3]{gliese1991}. We selected the closest candidates for a 
spectroscopic follow-up performed at La Silla Observatory (Chile). 

Our sample 
of nearby candidates is described in \S~\ref{sample}. \S~\ref{spectro} 
describes the spectroscopic observations. In \S~\ref{classification} we review
the spectroscopic classification schemes and compute the spectral type of the
stars. Computation of distances is detailed in \S~\ref{distances}. 
In \S~\ref{halpha} we discuss the chromospheric activity of the stars in our
sample and the conclusions are given in \S~\ref{conclusion}.

\section{Nearby candidates sample}
\label{sample}

Due to their small distance, most of the nearby stars have high proper motions.
However, many stars in high proper motion catalogues have no distance
determined yet. Note that a high proper motion does not imply a short distance but may
also indicate a high space velocity with respect to the Sun. 

As a contribution towards in improving the completeness of the solar neighbourhood
census, we made a systematic search for nearby stars among two new high proper
motion catalogues. The cross-identification of those with the DENIS database
allowed us to determine the distances of the stars from NIR photometry, an 
appropriate wavelength range for M-type dwarfs.

In \cite{reyle2002}, we present preliminary distance estimates for 301 APM high 
proper motion stars \citep{scholz2000}. 15 stars were found to be within the 
25~pc limit of the \textit{Catalogue of Nearby Stars} 
\citep[CNS3,][]{gliese1991}. In \cite{reyle2004}, we used the 
Liverpool-Edinburgh high proper motion survey \citep{pokorny2003} and found 
100 stars that probably lie within 25 parsecs from the Sun. They are mainly M 
dwarf candidates, with the exception of few white dwarf candidates. 

\begin{table*}
\label{tab2}
\caption{Nearby APM star candidates observed at ESO/NTT. $B_J$ and $R$ are from the APM high proper motion catalogue, obtained on digitized plates. $I$ and $I-J$ are from DENIS.}
\begin{tabular}{lcccccccp{.9cm}lp{.7cm}}
\hline
name &$\alpha_{J2000}$ &$\delta_{J2000}$ &ESO    &$\mu_x$ &$\mu_y$                  &$B_J$ &$R$ &$I$ &$I-J$ &$d_{\mbox{phot}}$ \\
     &                 &                 &epoch  &\multicolumn{2}{c}{(''yr$^{-1}$)} &              &          &\multicolumn{2}{c}{\hspace{-1.5cm}DENIS} &(pc)\\
\hline
LP 291-115        &00 33 13.26 &-47 33 17.8 &1993.7&  0.26&  0.15 &16.45 &14.20 &12.26 &1.79 &15.7       \\
APMPM J0103-3738  &01 02 49.77 &-37 37 46.8 &1990.7&  1.44&  0.30 &19.73 &17.30 &13.83 &2.67 &12.2       \\
APMPM J0145-3205  &01 45 11.27 &-32 05 09.6 &1991.9&  0.58&  0.18 &13.83 &12.14 &11.36 &1.52 &18.4       \\
LP 940-20         &01 49 42.37 &-33 19 21.5 &1991.9&  0.38&  0.11 &16.60 &14.36 &12.65 &1.76 &19.6       \\
LHS 5045     	  &01 52 52.07 &-48 05 39.2 &1988.9& -0.55& -0.20 &14.52 &12.40 &10.77 &1.60 &11.0       \\
LP 225-57         &02 34 20.95 &-53 05 35.4 &1994.0&  0.25& -0.34 &12.88 &10.78 &9.79  &1.49 &9.6        \\
L 127-33          &02 40 36.99 &-60 44 48.5 &1991.9&  0.19&  0.29 &15.35 &13.85 &12.34 &1.13 &116.7/30.4$^a$\\
APMPM J0255-5140  &02 55 13.92 &-51 40 23.1 &1992.8&  0.59&  0.23 &15.03 &12.68 &11.56 &1.67 &13.7       \\
LP 831-45         &03 14 17.96 &-23 09 31.2 &1993.0&  0.35&  0.19 &13.29 &11.46 &9.90  &1.44 &11.7       \\
LHS 5090          &04 04 31.58 &-62 59 10.4 &1990.0&  0.25& -0.46 &16.34 &14.49 &12.85 &1.25 &103.6/25.5$^a$\\
LHS 1656          &04 18 50.80 &-57 14 06.0 &1995.0&  0.27&  0.75 &12.97 &11.02 &10.75 &1.22 &42.5/10.7$^a$  \\
APMPM J0541-5349  &05 41 27.18 &-53 49 17.9 &1991.1&  0.10&  0.37 &13.61 &11.82 &11.77 &1.13 &90.3/23.6$^a$  \\
LP 904-51         &10 41 43.77 &-31 11 52.7 &1993.1&  0.37& -0.25 &16.80 &15.10 &12.84 &1.83 &19.3       \\
APMPM J1107-2202  &11 06 45.50 &-22 02 16.2 &1992.2&  0.34& -0.27 &13.92 &11.87 &11.48 &1.56 &16.8       \\
APMPM J1932-4834  &19 31 53.09 &-48 33 50.5 &1991.6&  0.01& -0.37 &16.21 &14.15 &12.39 &1.85 &15.4       \\
APMPM J2101-4115  &21 01 03.44 &-41 14 31.8 &1991.7&  0.36& -0.25 &15.40 &13.46 &11.50 &1.55 &17.5       \\
APMPM J2101-4907  &21 01 07.50 &-49 07 23.7 &1992.6& -0.31& -0.15 &12.01 &11.25 &10.52 &1.43 &16.5       \\
LTT 8708 A/B      &21 49 11.00 &-41 33 28.6 &1990.7&  0.30& -0.17 &12.65 &10.06 &9.29  &1.48 &7.8        \\
LHS 3800          &22 23 08.90 &-43 27 35.3 &1990.8&  0.79& -0.37 &15.39 &13.85 &12.23 &1.33 &15.7       \\
LHS 3842          &22 40 57.59 &-45 43 23.1 &1992.6&  0.33& -0.30 &14.39 &12.35 &11.30 &1.55 &16.0       \\
APMPM J2330-4737  &23 30 16.48 &-47 36 38.2 &1992.9& -0.59& -0.95 &19.13 &17.19 &13.70 &2.38 &15.0       \\
APMPM J2352-3609$^b$  &23 52 27.35 &-36 09 12.2 &1996.6&  0.31& -0.37 &18.55 &16.60 &$J$=13.28 &$J-K_S$=1.10 &23.3   \\
\hline\\
\end{tabular}

$^a$ Photometric distance estimate assuming that the star belongs to the disc/to the thick disc, obtained using
the ($M_J$,$I-J$) theoretical relation from \cite{baraffe1998} with a metallicity of 0 dex/-0.8 dex (see 
\cite{reyle2002}).\\
$^b$ $I$ drop-out star. The DENIS magnitude is in the $J$-band, the colour is $J-K_S$. The photometric distance has
been computed using the ($M_J$,$J-K_S$) theoretical relation from \cite{baraffe1998}.
\end{table*}

\begin{table*}
\label{tab2}
\caption{Nearby stars candidates found in the LEHPMS observed at ESO/NTT. $B_J$ and $R$ are from the LEHPMS high proper motion catalogue, obtained on digitized plates. $I$ and $I-J$ are from DENIS.}
\begin{tabular}{lcccccccccc}
\hline
name &$\alpha_{J2000}$ &$\delta_{J2000}$ &ESO      &$\mu_{\alpha}$ &$\mu_{\delta}$    &$B_J$ &$R$ &$I$ &($I-J$)      &$d_{\mbox{phot}}$ \\
     &                 &                 &epoch    &\multicolumn{2}{c}{(''yr$^{-1}$)} &      &          &\multicolumn{2}{c}{DENIS} &(pc)      \\
\hline
LEHPMS J0029-2733 &00 28 54.55 &-27 33 34.3& 1984.8&  0.23&  0.03& 16.65& 14.63& 12.47& 1.67& 19.9      \\
L 170-14A 	  &00 29 50.33 &-54 41 32.5& 1988.7& -0.27& -0.22& 14.51& 12.89& 10.99& 1.64& 14.6      \\
LEHPMS J0102-5524 &01 01 57.31 &-55 23 47.1& 1981.8&  0.28&  0.07& 15.34& 13.20& 11.24& 1.54& 18.5      \\
LHS 1201 	  &01 08 46.93 &-37 10 16.8& 1989.7& -0.56& -0.45& 17.29& 14.92& 12.72& 1.92& 17.9      \\
LHS 1249$^a$ &01 24 03.54 &-42 40 30.0& 1983.8&  0.31& -0.54& 15.62& 14.81& 14.46& 0.25& 16.9     \\
LEHPMS J0141-5544$^a$ &01 40 31.11 &-55 43 39.4& 1984.9&  0.21& -0.01& 15.12& 15.28& 14.97& -0.63& $--$ \\
LHS 1293 	  &01 45 11.07 &-32 05 09.7& 1985.9&  0.67&  0.22& 15.37& 13.25& 11.20& 1.51& 18.0   \\
LEHPMS J0146-5230 &01 45 30.37 &-52 30 20.1& 1984.9&  0.18&  0.03& 15.26& 13.86& 12.17& 1.71& 17.5   \\
LEHPMS J0146-5340 &01 46 29.07 &-53 39 30.9& 1984.9&  0.18& -0.11& 15.52& 13.18& 10.90& 1.59& 12.7   \\
LHS 1367 	  &02 15 07.38 &-30 39 57.2& 1988.8&  0.77& -0.34& 20.23& 17.34& 14.03& 2.46& 17.4   \\
NLTT 829-41 	  &02 16 21.55 &-22 00 52.0& 1984.8& -0.12&  0.22& 18.14& 15.55& 12.74& 1.96& 19.6   \\
NLTT 941-57 	  &02 22 18.07 &-36 51 52.8& 1989.8&  0.29&  0.10& 17.45& 15.16& 12.57& 1.93& 18.2   \\
NLTT 8435$^a$     &02 35 21.91 &-24 00 38.3& 1985.8& -0.11& -0.63& 16.58& 15.36& 14.92& 0.56& 14.3   \\
LP 942-107 	  &03 05 10.79 &-34 05 23.1& 1983.8&  0.37& -0.05& 14.41& 11.97&  9.72& 1.55& 14.8   \\
LEHPMS J0309-4925 &03 08 59.98 &-49 24 53.5& 1981.8&  0.16&  0.10& 16.19& 13.67& 11.45& 1.69& 17.5   \\
NLTT 994-114 	  &03 09 21.59 &-39 11 02.8& 1981.9&  0.38& -0.01& 13.56& 11.25&  9.99& 1.42& 15.9   \\
NLTT 772-8 	  &03 09 50.86 &-19 06 46.9& 1986.9&  0.41& -0.09& 15.41& 12.85& 10.94& 1.56& 17.4   \\
LHS 1524 	  &03 17 17.57 &-19 40 14.7& 1986.9&  0.53& -0.25& 16.57& 14.23& 11.86& 1.78& 19.3   \\
LEHPMS J0321-6352 &03 20 51.84 &-63 51 47.9& 1984.7&  0.04& -0.31& 14.29& 11.77&  9.49& 1.48& 14.0   \\
LTT 1732 	  &03 38 55.65 &-52 34 15.0& 1981.8&  0.15&  0.21& 15.74& 13.18& 11.15& 1.58& 14.6   \\
LEHPMS J0428-6209 &04 28 05.44 &-62 09 28.6& 1989.8&  0.21&  0.32& 14.15& 11.76&  9.96& 1.50& 11.9   \\
LEHPMS J0503-5353 &05 03 24.90 &-53 53 20.9& 1982.0&  0.48&  0.10& 15.61& 13.14& 11.02& 1.64& 16.6   \\
NLTT 83-11 	  &22 10 10.48 &-70 10 06.8& 1984.6&  0.08& -0.27& 14.03& 11.78& 10.53& 1.46& 17.9   \\
LEHPMS J2210-7010 &22 10 19.97 &-70 10 04.4& 1984.6&  0.08& -0.26& 14.88& 12.66& 11.17& 1.67& 14.2   \\
LEHPMS J2230-5345 &22 30 09.59 &-53 44 44.8& 1985.7& -0.03& -0.76& 15.81& 13.39& 11.19& 1.68& 12.0   \\
LEHPMS J2234-6108 &22 34 04.12 &-61 07 40.4& 1985.5&  0.26& -0.05& 16.36& 13.84& 10.88& 1.65& 15.9   \\
NLTT 1033-31      &22 35 04.52 &-42 17 45.3& 1981.8&  0.27& -0.16& 14.40& 11.80& 10.34& 1.40& 18.3   \\
NLTT 166-3 	  &22 41 59.52 &-59 15 12.2& 1986.8&  0.34&  0.00& 15.34& 13.07& 11.25& 1.71& 12.5   \\
LEHPMS J2248-4715$^a$ &22 48 08.68 &-47 14 45.3& 1983.8&  0.05& -0.29& 17.77& 15.84& 15.26&  0.50& 18 \\
NLTT 877-72  	  &23 09 59.91 &-21 11 43.1& 1984.8&  0.34&  0.01& 13.67& 10.89&  9.82& 1.36& 19.1    \\
LEHPMS J2325-6740 &23 25 24.54 &-67 40 05.6& 1984.8&  0.24& -0.15& 15.60& 13.22& 11.09& 1.61& 16.5    \\
LP 878-73 	  &23 40 47.99 &-22 25 27.9& 1986.6& -0.37& -0.25& 16.36& 14.10& 12.07& 1.77& 19.2    \\
LHS 73$^b$ &23 43 15.49 &-24 10 47.2& 1986.8&  1.37& -2.20& 14.03& 11.72& 11.17& 1.09& 12.3 \\
LTT 9783 	  &23 53 08.12 &-42 32 04.5& 1989.8&  0.18&  0.10& 13.93& 11.24&  9.86& 1.44& 15.9     \\
NLTT 987-47       &23 53 40.74 &-35 59 07.6& 1988.8&  0.30&  0.16& 14.77& 12.46& 10.68& 1.43& 19.1     \\
LEHPMS J2354-3634$^a$ &23 54 18.80 &-36 33 47.3& 1988.8&  0.05& -0.68& 15.60& 15.71& 15.34& -0.50& $--$ \\
LEHPMS J2359-5400 &23 58 49.37 &-54 00 13.2& 1989.6& -0.15& -0.16& 15.64& 13.05& 11.57& 1.56& 19.6     \\
\hline\\
\end{tabular}

$^a$ White dwarf candidates. The photometric distance has been computed with ($M_J$,$I-J$) theoretical 
relation from \cite{bergeron1995} for 0.6\msun white dwarfs. No distances were determined for the 
bluest candidates as they are beyond the colour limit of the theoretical relation.\\
$^b$ Subdwarf candidate. The photometric distance has been obtained using
the ($M_J$,$I-J$) theoretical relation from \cite{baraffe1998} with a metallicity of -1.8 dex.
\end{table*}

\begin{figure}
\includegraphics[width=6.cm,clip=,angle=-90]{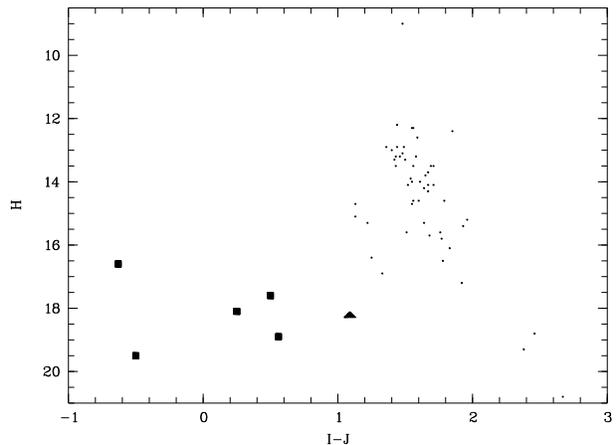}
   \caption{($H$,$I-J$) diagram for the high proper motion stars cross-identified with 
DENIS. Dots: M-dwarfs. Squares: white dwarf candidates. Triangle: subdwarf candidate.}
   \label{mured}
\end{figure}

Given the large uncertainties on the photometric distances (up to 45\% in some 
cases), follow-up observations of these candidates were needed. We selected 54 
stars, with photometric distances smaller than 20 pc, for spectroscopic 
follow-up. Using the reduced proper motion versus colour diagram, we showed 
that these objects are likely M dwarfs, except 5 white dwarf candidates. We also observed 4 
stars that could be either disc or thick disc M dwarfs, for which we determined 
two photometric distances with two metallicity hypotheses. They are nearby 
candidates if they belong to the thick disc. Finally, we 
obtained spectra for one star that may 
be a nearby subdwarf and one star detected in the J and K bands, but 
not in the I band.
Optical and near-IR DENIS photometry,
proper motion, as well as photometric distance determination are given in
Tables~2 and 3 for the APM and LEHPMS nearby candidates, respectively.
Figure~\ref{mured} shows their reduced proper motion versus colour diagram,
where the reduced proper motion is $H = I + 5 \mbox{log}\mu +5$.

\section{Spectroscopic observations}
\label{spectro}

In order to insure the spectral types and therefore distances, we carried out 
spectroscopic observations on the 3.6m New Technology Telescope (NTT) at La 
Silla Observatory (ESO, Chile) in November 2003. Optical low-resolution 
spectra were obtained in the Red Imaging and Low-dispersion spectroscopy 
(RILD) observing mode with the EMMI instrument.
The spectral dispersion of the grism we used is 0.28 nm/pix, with a wavelength
range 385-950~nm. Except for the white dwarf candidates, we used an
order blocking filter to avoid the second order overlap that occurs beyond
800 nm. Thus the effective wavelength coverage ranges from 520 to 950 nm. 
The slit was 1 arcsec wide and the resulting resolution was 10.4~\AA~. The 
seeing varied from 0.5 to 1.5 arcsec. Exposure
time ranged from 15~s for the brightest to 120~s for the faintest dwarf ($I$ = 15.3). 
The reduction of the spectra was done using the context 
\textit{long} of MIDAS. Fluxes were calibrated with the spectrophotometric standards 
LTT~2415 and Feige~110.

\begin{figure*}
\includegraphics[width=21.cm,clip=,angle=-90]{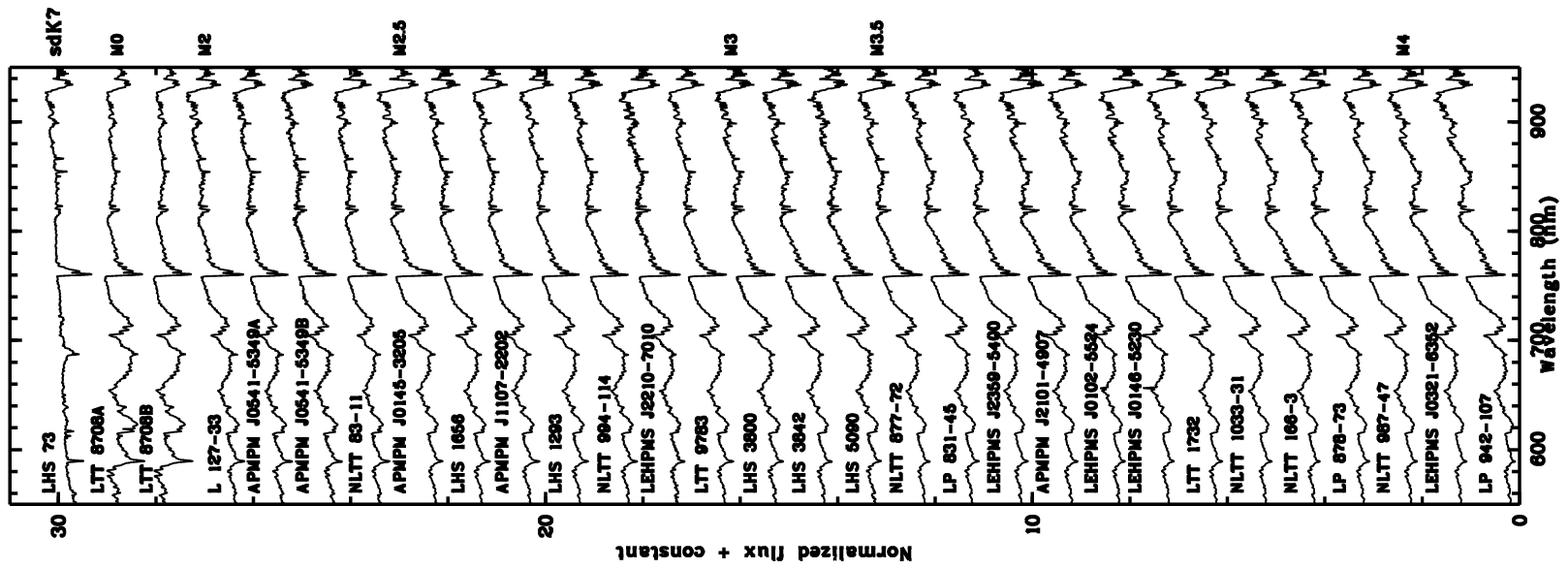}
\hspace*{-0cm}\includegraphics[width=21.cm,clip=,angle=-90]{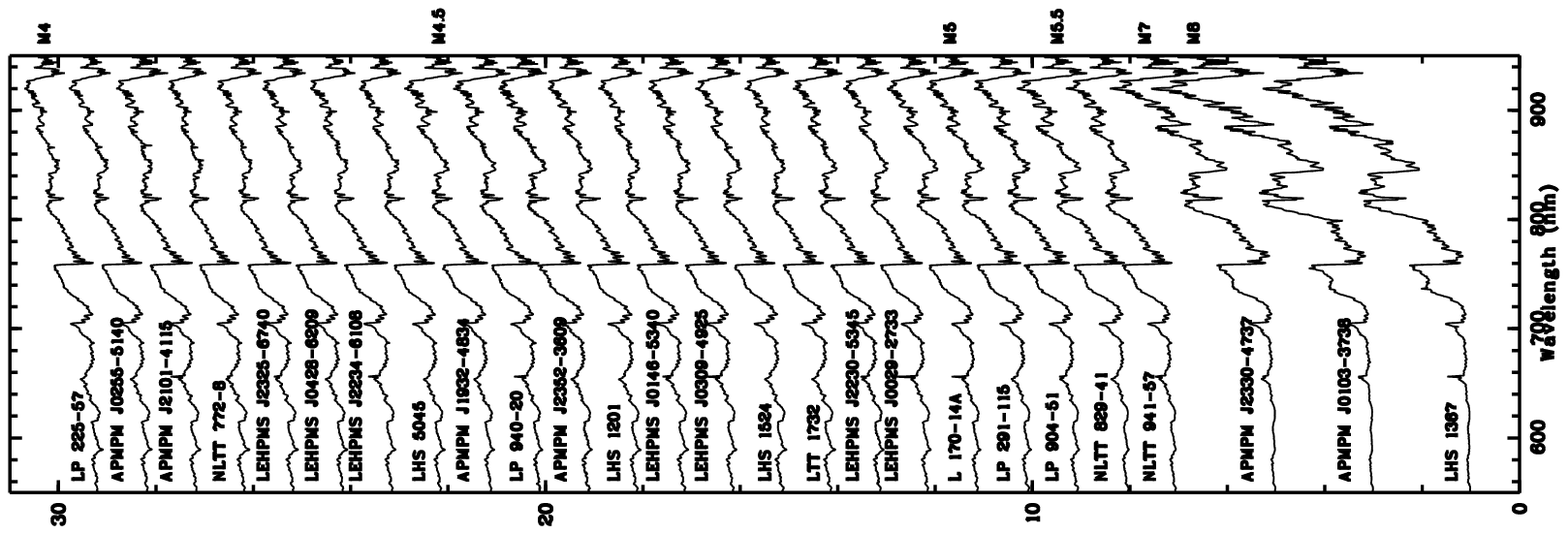}
   \caption{NTT spectra of K and M dwarfs, from spectral types K7 to M4 on 
the left panel, and from M4 to M8 on the right panel.}
   \label{fig1}
\end{figure*}

We obtained spectra for 59 nearby candidates. They are shown in Figure~\ref{fig1}
and ~\ref{fig2}. 
In addition, we observed 
two radial velocity standard M-stars \citep{Nidever2002} with 
spectral type sdM0 and M1, 
12 comparison stars with known spectral types from M0 to M6, 
found in the NASA NStars\footnote{http://nstars.arc.nasa.gov}, 
the ARICNS\footnote{http://www.ari.uni-heidelberg.de/aricns/} 
and the SIMBAD\footnote{http://cdsweb.u-strasbg.fr/Simbad} databases, 
and one M9 star \citep{delfosse2001}.
Discrepancies are found in the spectral types given by the different 
databases (see Table~4), probably due to the fact that different spectral 
classification systems are used. 
These templates, and our finally adopted types are in  currently 
accepted system of \cite{kirkpatrick1991}. Deviations between our adopted
sequence and other published spectral types are usually 0.5 to 1 subtype and
discrepancies mostly occur at earlier types (M0-M3).

\begin{table}
\label{tab4}
\caption{Comparison stars for spectroscopic classification. The spectral types found in the literature are
given, as well as the final adopted spectral types.}
\begin{tabular}{p{1.65cm}p{.6cm}p{.8cm}p{.8cm}p{1.8cm}p{.8cm}}
\hline
name &NStars	&ARICNS &Simbad &other &adopted \\
\hline
LHS 29           &-- &M1	&M1	&sdM1.0$^a$/sdM0$^b$ &sdM0\\ 
Gl 143.1         &M0 &M0 &K7.0 &-- &M0\\
LHS 141          &M0.5 &M1.0 &K7 &-- &M0  \\
LHS 3833         &-- &M0 &M0 &-- &M0.5\\ 
LHS 1827         &M1 &M1 &M1/M2 &-- &M1  \\ 
LHS 65           &M3 &M2 &M1 &-- &M1.5\\
LHS 14           &M2.5 &M1.5 &M1.5 &-- &M1.5\\
LHS 1208         &M2 &-- &M2 &-- &M2.5\\
LHS 502          &M4 &M3 &M2.5 &-- &M3  \\
LHS 183          &M3.5 &M4 &M3.5 &--  &M3.5\\
LHS 138          &M4.5 &M4.5 &M4.5 &--  &M4.5\\
LHS 168          &M5.0 &M5.0 &M5.0 &--  &M5  \\
LHS 546          &M5.5 &M5.5 &M5.5 &-- &M5.5\\
LHS 1326         &M6 &M6 &M6 &--  &M6  \\             
D\sc{enis} J1048 &-- &-- &M8 &M9$^c$   &M9  \\
\hline\\
\end{tabular}

$^a$ \citet{gizis1997} \\
$^b$ \citet{Nidever2002} \\
$^c$ \citet{delfosse2001} 
\end{table}

\section{Spectral classification}
\label{classification}

\subsection{M dwarfs}
In the past years, several classification schemes have been proposed for 
cool dwarfs. First, the least-squares minimization technique by 
\cite{kirkpatrick1991} has the advantage of assigning types based on both 
molecular or atomic features and the slope of the spectrum. Recently, 
\cite{henry2002} developed a similar but more accurate method that 
matches a target spectrum to one from a database of standard spectra 
(software program called ALLSTAR). \cite{scholz2005} used the minimisation
of the absolute differences between target and template spectra in their
low-resolution classification spectroscopy.

A different approach was taken by \cite{reid1995}, who measured the strengths
of the prominent TiO and CaH features of M dwarfs. They defined several
indices being the ratio between the flux within a given spectral feature and 
the flux in a nearby pseudo-continuum region. 

For the latest M dwarfs, \cite{kirkpatrick1995} defined an index based on the 
VO bandstrength, whereas \cite{kirkpatrick1999} and \cite{martin1999} proposed 
spectral indices for classification down to the L type where TiO and VO bands 
fade due to the condensation of Ti and V to dust. The pseudo-continuum spectral
ratios (namely PC3) defined by \cite{martin1999} covers a narrower wavelength 
range than for the computation of indices, thus it is less sensitive to possible 
wavelength shifts.

\cite{lepine2003a} pointed out that a 
unified scheme for the classification of M dwarfs is needed and they defined an 
enhanced classification based on a set of existing \citep{reid1995,hawley2002}
and new spectral indices that measure the most important features in the optical spectral range. 

Furthermore, the classification procedure of \cite{gizis1997} allows us to distinguish
potential subdwarfs (sdM) or extreme subdwarfs (esdM) from the M dwarfs. 
Accordingly, \cite{lepine2003b} used a CaH2+CaH3 versus TiO5 diagram to 
discriminate between the three different object classes. This diagram was 
also used by \cite{lepine2004} and \cite{scholz2004a} when reporting the discovery
of an esdM6.5 and an sdM9.5 star respectively. Figure~\ref{fig3} shows the locus
of our candidates (filled circles) in the CaH2+CaH3 versus TiO5 diagram superimposed 
with the data used in the same diagram by \cite{lepine2003b} and newly discovered
objects by \cite{lepine2004,scholz2004a,scholz2004b,farihi2005}. 

\begin{figure}
\includegraphics[width=14cm,clip=,angle=-90]{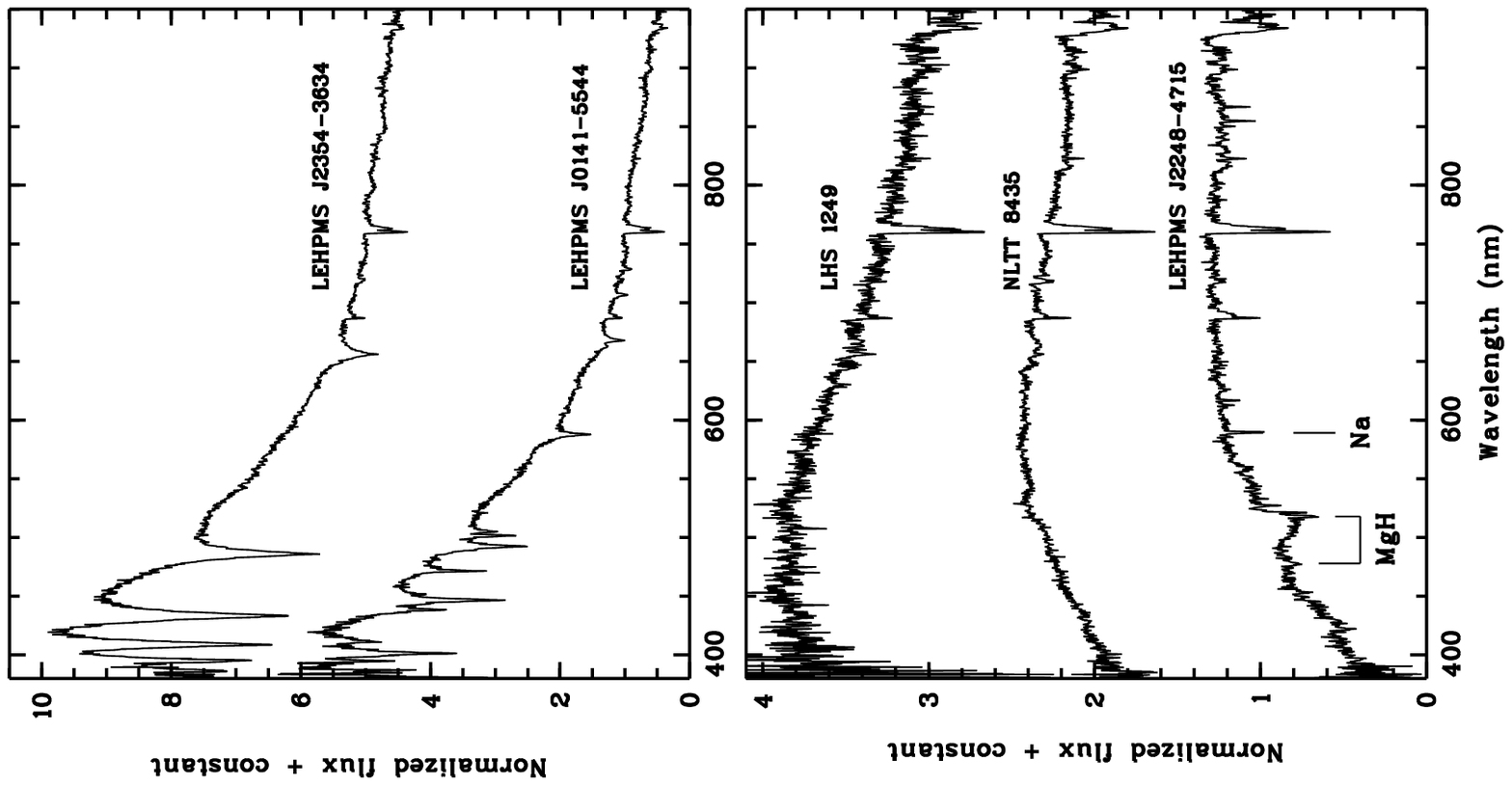}
   \caption{NTT spectra of the white dwarf candidates. Two hot white dwarfs are
shown in the top panel, whereas two featureless cool white dwarfs and a K subdwarf
(bottom) are shown in the lower panel.}
   \label{fig2}
\end{figure}

\begin{figure}
\centering
\includegraphics[width=8cm,clip=,angle=-90]{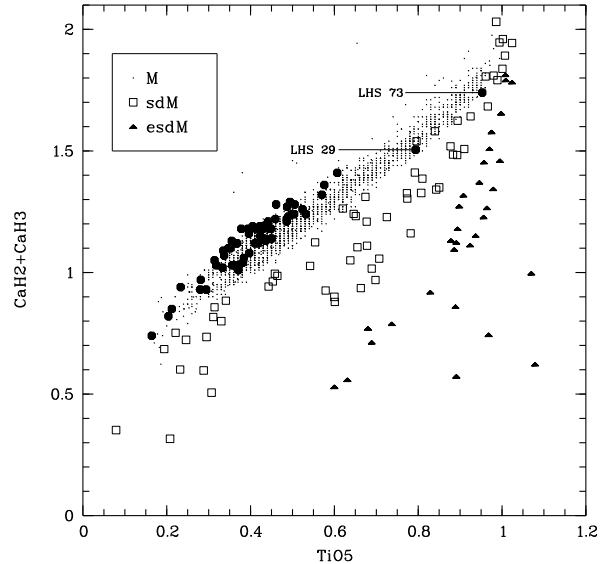}
   \caption{Relative strengths of the CaH and TiO molecular bands for spectroscopically
identified M dwarfs, subdwarfs (sdM) and extreme subdwarfs (esdM) according to
the classification scheme of \citet{gizis1997}. Our nearby candidates are shown with 
filled circles.}
   \label{fig3}
\end{figure}

Most of our candidates lie in the M-dwarfs region, including the 4 stars for which
alternative distances were computed assuming that they belong to the thick disc
(see last column in Table~2). 
Thus these stars are more distant disc stars instead of nearby low-metallicity dwarfs.

Spectral types are obtained  by visual comparison with our spectral templates of comparison
stars, observed together with our target stars (Table~4). For comparison, we also derive spectral types, using the classification scheme based on the TiO and CaH bandstrengths. We also measured the pseudo-continuum spectral ratio PC3, useful for late dwarfs, in particular for M7 and M8 stars that are missing in our  comparison sequence. The results are given in Table~5. The uncertainty of our spectral type determination is 0.5 subclass. 

\begin{figure}
\centering
\includegraphics[width=7cm,clip=,angle=-90]{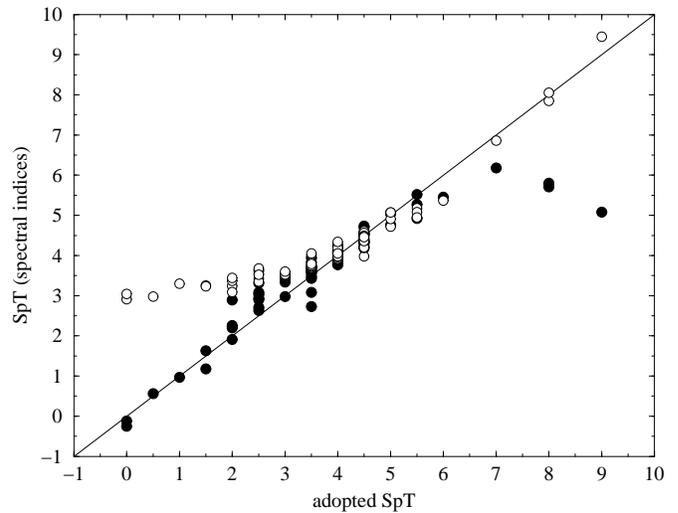}
   \caption{Spectral type obtained from the TiO5 (filled circles) and PC3 (open circles) indices versus 
the adopted spectral type checked against the spectra of comparison stars.}
   \label{fig4}
\end{figure}

Figure~\ref{fig4} shows the spectral type determined from the TiO5 and PC3 indices
versus our adopted spectral type obtained from the comparison with templates. As
already pointed out by the different authors who defined the indices, the TiO5 index starts 
to deviate from our spectral sequence for spectral types later than M6, whereas the PC3 index is not 
reliable for stars earlier than M3. Both indices are well in agreement in the overlapping region
M3-M6.

\subsection{Subdwarfs}

The sdM0 star
LHS~29 and LHS~73 are also marked in Figure~\ref{fig3}. The latter could also be a subdwarf
given its location in the reduced proper motion versus colour diagram. Its
position in the CaH2+CaH3 versus TiO5 diagram is compatible with the position
of other known subdwarfs, although dwarfs, subdwarfs and extreme subdwarfs 
are not well separated in this part of the diagram.

For this star, LHS~73, we determined the radial
velocity. We compared the position of the potassium line at 769.9 nm, the iron line at
838.8 nm, and the calcium triplet at 849.8 nm, 854.2 nm and 866.2 nm with the position 
of the same lines in the spectrum of the standard star LHS 1827 with known radial velocity
(47.2 km s$^{-1}$). We obtained a radial velocity of 100.6 km s$^{-1}$, with an accuracy of
15 km s$^{-1}$. We then computed the galactocentric velocities and found ($U,V,W$) =
(-0.4 $\pm$ 3.8,-207.0 $\pm$ 23.3,-141.4 $\pm$ 15.2) km s$^{-1}$, compatible with the space motion 
of a star belonging to an old population, thick disc or halo.
Thus, following the procedure of \cite{gizis1997}, we derived a sdK7 spectral type.

\begin{table*}
\label{tab5}
\caption{Spectroscopic indices, spectral type, absolute magnitude $M_J$ and spectroscopic distance 
of the red dwarf candidates observed at NTT. The additional letter ``e'' in the spectral type indicates a star showing $H_\alpha$ emission.}
\scriptsize{
\begin{tabular}{lcccccccccl}
\hline
name                 &TiO5  &CaH1  &CaH2  &CaH3  &PC3   &SpT &$J$ &$M_J$ &$d_{\mbox{spectro}}$  &Comments\\ 
                     &      &      &      &      &      &    &    &      &(pc)                  &\\ 
\hline
LP 291-115           &0.281 &0.784 &0.334 &0.646 &1.268 &M5.0 &10.47 & 9.72  &14.2$\pm$2.0      &\\ 
APMPM J0103-3738     &0.212 &0.891 &0.265 &0.590 &1.801 &M8.0e &11.16 &11.15  &10.0$\pm$0.8     &\\ 
APMPM J0145-3205     &0.504 &0.803 &0.497 &0.754 &1.055 &M2.5 & 9.84 & 7.14  &34.7$\pm$2.8      &\\ 
LP 940-20        &0.352 &0.766 &0.418 &0.696 &1.188 &M4.5e &10.89 & 8.84  &25.7$\pm$8.4 &M4.5 at $d$ = 31.4 $\pm$ 5.3 pc$^b$\\ 
LHS 5045     	     &0.365 &0.840 &0.417 &0.718 &1.171 &M4.0 & 9.17 & 8.63  &12.9$\pm$4.4      &\\ 
LP 225-57            &0.401 &0.854 &0.452 &0.736 &1.123 &M4.0 & 8.23 & 8.63  & 8.3$\pm$2.8      &\\ 
L 127-33             &0.570 &0.801 &0.558 &0.778 &1.031 &M2.0 &11.21 & 6.97  &70.6$\pm$5.3      &\\ 
APMPM J0255-5140     &0.394 &0.858 &0.441 &0.745 &1.132 &M4.0 & 9.89 & 8.63  &17.9$\pm$6.1      &\\ 
LP 831-45            &0.444 &0.808 &0.464 &0.727 &1.102 &M3.5 & 7.62 & 8.41  &10.4$\pm$3.5      &\\ 
LHS 5090             &0.450 &0.782 &0.434 &0.713 &1.067 &M3.0 &11.60 & 7.32  &71.8$\pm$6.2      &\\ 
LHS 1656             &0.500 &0.811 &0.501 &0.749 &1.073 &M2.5 & 9.53 & 7.15  &30.0$\pm$2.4      &\\ 
APMPM J0541-5349A    &0.576 &0.794 &0.575 &0.798 &1.048 &M2.0 &10.58 & 6.97  &52.7$\pm$4.0      &\\ 
APMPM J0541-5349B    &0.505 &0.780 &0.526 &0.762 &1.058 &M2.0 &10.64 & 6.97  &54.2$\pm$4.1      &\\ 
LP 904-51            &0.280 &0.743 &0.333 &0.606 &1.292 &M5.0e &11.01 & 9.72  &18.1$\pm$2.6     &\\ 
APMPM J1107-2202     &0.532 &0.779 &0.505 &0.749 &1.080 &M2.5 & 9.92 & 7.14  &36.0$\pm$2.9      &\\ 
APMPM J1932-4834     &0.335 &0.846 &0.397 &0.707 &1.162 &M4.5 &10.54 & 8.84  &21.9$\pm$7.2      &\\ 
APMPM J2101-4115     &0.381 &0.755 &0.396 &0.652 &1.161 &M4.0e & 9.95 & 8.63  &18.4$\pm$6.3     &\\ 
APMPM J2101-4907     &0.432 &0.767 &0.443 &0.701 &1.135 &M3.5 & 9.09 & 8.41  &13.7$\pm$4.6      &\\ 
LTT 8708 A           &0.487 &0.766 &0.537 &0.749 &0.886 &M0.0 & 7.75$^a$ & 6.33  &27.2$\pm$1.8  &\\ 
LTT 8708 B           &0.494 &0.780 &0.542 &0.755 &0.879 &M0.0 & 7.75$^a$ & 6.33  &27.2$\pm$1.8  &\\ 
LHS 3800             &0.442 &0.785 &0.484 &0.737 &1.069 &M3.0 &10.90 & 7.32  &52.0$\pm$4.6      &\\ 
LHS 3842             &0.459 &0.795 &0.487 &0.744 &1.079 &M3.0 & 9.75 & 7.32  &30.6$\pm$2.6      &\\ 
APMPM J2330-4737     &0.164 &0.769 &0.233 &0.510 &1.596 &M7.0e &11.32 &10.73  &13.1$\pm$1.5 &M6 at $d$ = 15.8 $\pm$ 1.9 pc$^c$\\ 
APMPM J2352-3609     &0.314 &0.856 &0.379 &0.688 &1.203 &M4.5 &12.18 & 8.84  &46.6$\pm$15.3     &\\ 
LEHPMS J0029-2733    &0.368 &0.820 &0.421 &0.703 &1.162 &M4.5e &10.80 & 8.84  &24.7$\pm$8.1     &\\ 
L 170-14A 	     &0.347 &0.869 &0.409 &0.709 &1.188 &M4.5e & 9.35 & 8.84  &12.7$\pm$4.1     &\\ 
LEHPMS J0102-5524    &0.423 &0.809 &0.449 &0.711 &1.114 &M3.5e & 9.70 & 8.41  &18.2$\pm$6.2     &\\ 
LHS 1201 	     &0.318 &0.780 &0.370 &0.663 &1.215 &M4.5 &10.80 & 8.84  &24.7$\pm$8.1      &\\ 
LHS 1293 	     &0.490 &0.804 &0.497 &0.740 &1.084 &M2.5 & 9.69 & 7.14  &32.3$\pm$2.6      &\\ 
LEHPMS J0146-5230    &0.436 &0.803 &0.441 &0.692 &1.097 &M3.5e &10.46 & 8.41  &25.8$\pm$8.7     &\\ 
LEHPMS J0146-5340    &0.371 &0.761 &0.380 &0.637 &1.204 &M4.5e & 9.31 & 8.84  &12.4$\pm$4.1     &\\ 
LHS 1367 	     &0.204 &0.849 &0.262 &0.561 &1.846 &M8.0e &11.57 &11.15  &12.1$\pm$1.0     &\\ 
NLTT 829-41         &0.233 &0.757 &0.310 &0.631 &1.293 &M5.5 &10.78 &10.05  &14.0$\pm$1.7   &M5.5 at $d$ = 19.3 $\pm$ 1.6 pc$^b$\\ 
NLTT 941-57 	     &0.294 &0.776 &0.335 &0.603 &1.273 &M5.5e &10.64 &10.05  &13.1$\pm$1.6     &\\ 
LP 942-107 	     &0.405 &0.839 &0.450 &0.741 &1.128 &M4.0 & 9.63 & 8.63  &15.9$\pm$5.4      &M4 at $d$ = 17.9 pc$^d$\\ 
LEHPMS J0309-4925    &0.359 &0.770 &0.393 &0.641 &1.188 &M4.5e & 9.76 & 8.84  &15.3$\pm$5.0     &\\ 
NLTT 994-114 	     &0.486 &0.797 &0.485 &0.739 &1.090 &M2.5 & 9.00 & 7.14  &23.5$\pm$1.9      &\\ 
NLTT 772-8 	     &0.410 &0.774 &0.433 &0.696 &1.167 &M4.0 & 9.38 & 8.63  &14.2$\pm$4.8      &\\ 
LHS 1524 	     &0.337 &0.793 &0.390 &0.687 &1.185 &M4.5 &10.08 & 8.84  &17.7$\pm$5.8      &\\ 
LEHPMS J0321-6352    &0.414 &0.825 &0.451 &0.732 &1.132 &M4.0 & 9.13 & 8.63  &12.6$\pm$4.3      &\\ 
LTT 1732 	     &0.356 &0.743 &0.385 &0.659 &1.204 &M4.5 & 9.57 & 8.84  &14.0$\pm$4.6      &\\ 
LEHPMS J0428-6209    &0.449 &0.816 &0.460 &0.725 &1.111 &M3.5 & 8.94 & 8.41  &12.8$\pm$4.3      &\\ 
LEHPMS J0503-5353    &0.378 &0.814 &0.440 &0.746 &1.162 &M4.0 & 9.38 & 8.63  &14.2$\pm$4.8      &\\ 
NLTT 83-11 	     &0.607 &0.856 &0.609 &0.818 &1.011 &M2.0 & 9.07 & 6.97  &26.3$\pm$2.0      &\\ 
LEHPMS J2210-7010    &0.461 &0.802 &0.526 &0.764 &1.046 &M3.5 & 9.50 & 8.41  &16.6$\pm$5.6      &\\ 
LEHPMS J2230-5345    &0.333 &0.794 &0.371 &0.650 &1.200 &M4.5 & 9.51 & 8.84  &13.6$\pm$4.5      &\\ 
LEHPMS J2234-6108    &0.366 &0.745 &0.393 &0.649 &1.183 &M4.0e & 9.23 & 8.63  &13.2$\pm$4.5     &\\ 
NLTT 1033-31         &0.422 &0.836 &0.461 &0.733 &1.111 &M3.5 & 9.10 & 8.41  &13.8$\pm$4.7      &\\ 
NLTT 166-3 	     &0.396 &0.827 &0.443 &0.726 &1.096 &M3.5 & 9.54 & 8.41  &16.9$\pm$5.7      &\\ 
NLTT 877-72  	     &0.486 &0.796 &0.486 &0.742 &1.109 &M3.5 & 8.86 & 8.41  &12.4$\pm$4.2      &\\ 
LEHPMS J2325-6740    &0.397 &0.791 &0.416 &0.677 &1.142 &M4.0e & 9.48 & 8.63  &14.8$\pm$5.0     &\\ 
LP 878-73 	     &0.418 &0.783 &0.430 &0.698 &1.141 &M3.5 &10.30 & 8.41  &23.9$\pm$8.1      &\\ 
LHS 73    &0.960 &0.889 &0.847 &0.901 &0.997 &sdK7 &10.08 & 8.72 &18.7          \\ 
LTT 9783 	     &0.524 &0.808 &0.514 &0.757 &1.068 &M2.5 & 9.17 & 7.14  &25.5$\pm$2.0      &\\ 
NLTT 987-47          &0.430 &0.839 &0.459 &0.734 &1.100 &M3.5 & 9.25 & 8.41  &14.8$\pm$5.0      &\\ 
LEHPMS J2359-5400    &0.431 &0.836 &0.466 &0.735 &1.105 &M3.5 &10.01 & 8.41  &21.0$\pm$7.1      &\\ 
\hline\\
\end{tabular}
}

\small
$^a$ Total magnitude of the system
$^b$ \cite{reid2003b}
$^c$ \cite{lodieu2005}
$^d$ \cite{scholz2005}

\end{table*}

\subsection{White dwarfs}
\label{wd}
The spectra of the five white dwarf candidates are shown in Figure~\ref{fig2}.
The two bluest candidates shown on the top panel are hot white dwarfs. LEHPMS 
J2354-3634 is an early type DA white dwarf with broad Balmer lines and LEHPMS 
J0141-5544 is a helium rich DB white dwarf. For such blue objects, the atmosphere
models of \cite{bergeron1995} for DA and DB white dwarfs do not give an absolute
magnitude. To compute their distance, we used instead the relation from \cite{oppenheimer2001}

$$M_{B_J} = 12.73 + 2.58(B_J-R)$$

We obtained $d_{B_J}$ = 44.9 pc for LEHPMS J2354-3634.
Distances were also derived by \cite{pauli2003,salim2004} from spectroscopic observations. 
They found a distance of 62.6 pc and
59 $\pm$ 14 pc respectively. From spectral fitting, \cite{koester2001} found an
effective temperature around 14500~K.
For LEHPMS J0141-5544, we computed $d_{B_J}$ = 36.3 pc. 

On the bottom panel, LHS 1249 and NLTT 8435 are featureless DC white dwarfs,
although they could show weak $H_\alpha$ absorption, in particular  LHS 1249. Spectra
with better signal to noise would be required to assign a DC or DA class.
We derived $d_{B_J}$ = 14.4 pc
for LHS 1249 and $d_{B_J}$ = 13.8 pc for NLTT 8435.
For those objects, we also estimated the distance using the $M_J$ versus $I-J$ theoretical 
relation for white dwarfs of 0.6 \msun from \cite{bergeron1995} and obtained 16.9 pc for
LHS 1249 and 14.3 pc for NLTT 8435.
Following the conventional number 50~400/$T_{\mbox{eff}}$ to assign a spectral index, 
we performed
black-body fits to the spectra and obtained 6200 K for LHS 1249 and 
4800 K for NLTT 8435, with an accuracy of $\pm$ 100~K. The derived spectral types are 
DC8 and DC10, respectively. \cite{kawka2004} obtained spectroscopy of NLTT 8435 
and found it to be a DC white dwarf at 16 pc and with 
$T_{\mbox{eff}}$ = 5300 $\pm$ 200 K.

The lower spectrum shows clear indications of K dwarfs (MgH, Na features). 
LEHPMS J2248-4715 is not a nearby white dwarf but a distant K subdwarf. Following the
spectral classification scheme of \cite{gizis1997}, we measured the Ti05, CaH1, CaH2, and CaH3
indices and estimated a spectral type sdK5.

\section{Spectroscopic distances}
\label{distances}

\subsection{M dwarfs}

Spectral type versus absolute magnitudes relations can be used to derive the distances of our red 
dwarf targets. \cite{reid1995,henry1994,henry2002} computed relations between spectral type and
absolute magnitude in the $V$ band, \cite{lepine2003a} in the $R$, $B$, and $K_s$ bands.

Several calibrations have also been computed in the $J$-band:

\cite{dahn2002} determined trigonometric parallaxes for 28 late-type dwarfs and derived a
$M_J$(spectral type) relation valid for types later than M6.5.

\cite{hawley2002} used a well observed sample of nearby dwarfs with measured trigonometric
parallaxes, 2MASS $J$ magnitude and spectral type to calibrate a spectroscopic parallax 
relation $M_J$(spectral type). They adopted a fit with several line segments in the spectral 
type range from K7 to T8.

\cite{cruz2002} used a sample of 68 late-type dwarfs with well determined trigonometric 
parallaxes surveyed by the Palomar/Michigan State University (PMSU) survey to derive a
$M_J$(TiO5) relation. \cite{cruz2003} suplemented these data to fit a fourth-order
polynomial $M_J$(spectral type) relation, valid for types M6 to L8. 

\cite{scholz2005} 
provided $M_J$ for spectral types over the whole range from K0 to L8.

\cite{crifo2005} identified in the literature 27 stars with good trigonometric parallaxes 
and photometry and with known PC3 indices to derive a calibration of absolute magnitudes in
the near infrared as a function of the PC3 index. 

\begin{figure}
\centering
\includegraphics[width=7cm,clip=,angle=-90]{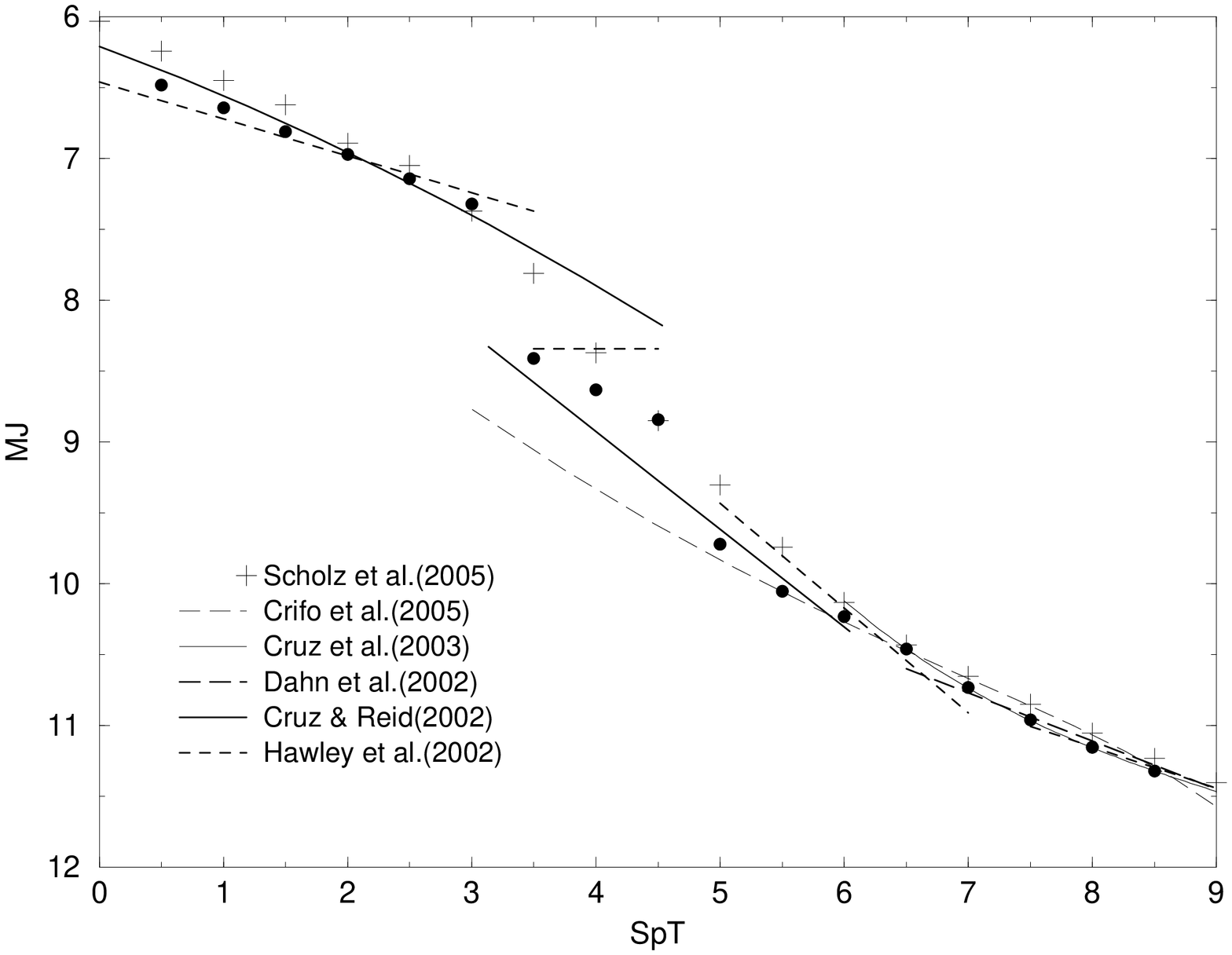}
   \caption{$M_J$ vs spectral type calibration relations from \citet{crifo2005} (long dashed line),
\citet{cruz2003} (thin line), \citet{hawley2002} (thick dashed line), \citet{cruz2002} (thick line), 
and \citet{dahn2002} (long dashed thick line). For the \citet{crifo2005} and \citet{cruz2002} 
relations, the PC3 and TiO5 indices had first to be transformed to spectral types, respectively. 
The calibration we adopted is a combination of these relations, and is shown by the filled 
circles. For comparison, the pluses show the values given by \citet{scholz2005}. Spectral types 
are from M0 to M9.}
   \label{fig5}
\end{figure}

Since the more accurate available photometry of our candidates is in the near infrared bands, 
we prefer to use spectral type versus near infrared absolute magnitudes relations. The 
available $M_J$ against spectral type calibration relations are shown in Figure~\ref{fig5}, in the
spectral type range where they are valid. We note that the absolute magnitudes $M_J$ given 
by most of calibration relations do not differ by more than the differences due to the
0.5 subclass uncertainty in the spectral type. 

For types earlier than M3.5, we computed the mean
of the $M_J$ obtained using \cite{hawley2002} and \cite{cruz2002} relations. For the stars falling in 
the region M3.5 to M4.5, where there is a clear discontinuity in the calibrating relations
\citep[see][for a discussion on this jump]{reid2002a}, 
we averaged the absolute magnitudes obtained from the upper \citep{cruz2002} and lower relations
\citep{cruz2002,crifo2005}, as well as the fixed value $M_J$ = 8.34 \citep{hawley2002}. For types M5 to M6, 
we used the relations by \cite{cruz2002} and \cite{crifo2005}. For later types, the relations are nearly 
equal. For practical reasons, we only used the relation by \cite{cruz2003} which is directly 
given as a function of spectral type. The filled circles in Figure~\ref{fig5} show the result of this
combination of existing calibrating relations. The values provided by \cite{scholz2005} are shown
for comparison (pluses). 

The resulting values of $M_J$ and spectroscopic 
distances of our candidates are given in Table~5. The uncertainties are also given, taking into 
account the 0.5 subclass uncertainty on the spectral type. They are less than 15\%. 
For the spectral types M3.5 to M4.5,  the uncertainties are computed given the upper and lower calibrations. The are larger, $\sim$ 34\%. For the brightest stars with 
$J < 9$, we used 2MASS $J$ magnitudes instead of DENIS due to saturation problems that may occur 
in the DENIS bands. 

Other determinations of spectral types and spectroscopic distances are given
in Table~5. They are in agreement with our values.

\subsection{Subdwarf}
\label{sd}

We computed the distance of the subdwarf LHS 73 assuming an absolute 
magnitude $M_J$ = 8.72, following \cite{reid2005}. The resulting distance is 18.7~pc.
LHS 73 has a common proper motion companion, LHS 72, with a separation of 94\arcsec.
LHS 72 has been spectroscopically classified as a subdwarf by \cite{rodgers1974}.
A quite uncertain trigonometric parallax, 37.6 $\pm$ 8.9 mas (21.5 pc $< d <$ 34.8 pc) 
can be found for this star in the Yale Trigonometric Parallaxes \citep{vanaltena1995}.

Most K/M subdwarfs with trigonometric parallaxes are listed in \cite{gizis1997}. Only a handful of 
these (LHS 20, LHS 29, LHS 35, LHS 44, LHS 55, LHS 103, LHS 3409) are
within 20 pc. Another pair of subdwarfs LHS 52/LHS 53, is slightly more distant ($d = 29.8 \pm 1.5$ pc).
To our knowledge, LHS 72/LHS 73 is the closest pair of subdwarfs.

\subsection{Binaries}
\label{binaires}
Besides LHS 73 which belongs to a common proper motion pair (\S~\ref{sd}), two objects in our sample are binaries. 

LTT 8708A/B is an already known system. We used the acquisition image,
obtained in the ESO $R$-band, to derive a separation of 8.1 pixels. With a scale of 
0.33\arcsec/pixel, this corresponds to an angular separation of 2.7\arcsec. The spectra of the 
two components show that they are both M0 stars, with absolute magnitudes $M_J$ = 6.33. With an 
observed total magnitude $J$ = 7.75, we derived a distance of 27.2 $\pm$ 1.8 pc. This result is 
in agreement with the trigonometric parallax derived by Hipparcos, although the uncertainty was 
very large (74.42 mas $\pm$ 41.23 or a distance between 8.6 pc and 30.1 pc). \cite{hawley1997}
found a smaller distance, $d$ = 17.5 pc, and a spectral type of M2.5.

\begin{figure}
\centering
\includegraphics[width=7cm,clip=,angle=0]{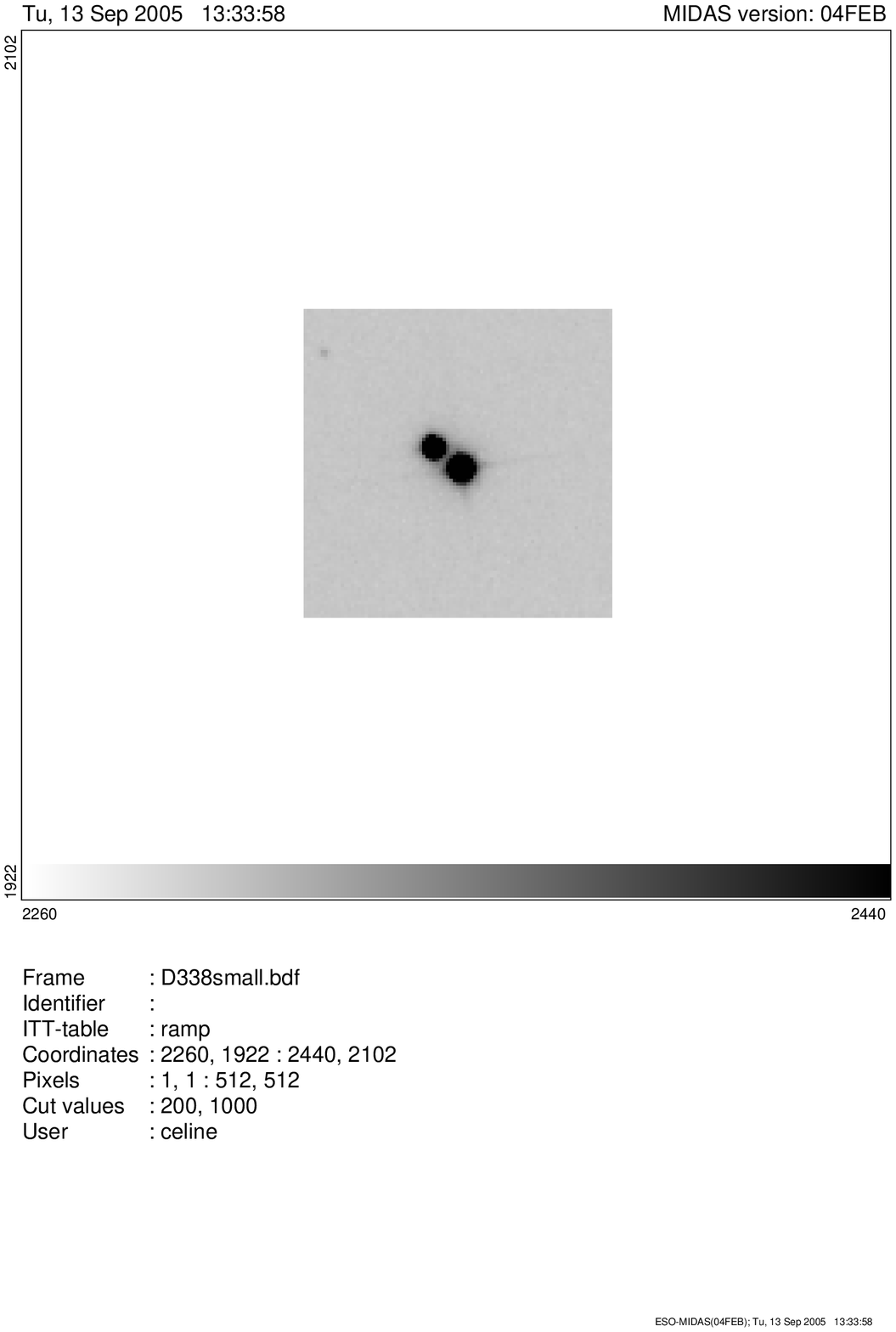}
   \caption{Part of the $R$-band NTT image of the binary APMPM J0541-5349. The size of the image 
is 30\arcsec $\times$ 30\arcsec. North is up, East is to the left.}
   \label{fig6}
\end{figure}

APMPM J0541-5349 is clearly seen as a binary on the acquisition image (Figure~\ref{fig6}). The 
separation between the components is 10.1 pixels, corresponding to 3.33\arcsec. The examination 
of earlier images such as 2MASS or DSS2 show that they are a common proper motion pair. Two 
objects are given in the 2MASS point source catalogue, with magnitudes $J$ = 10.58 and 10.64. 
Both components have a M2 spectral type, and an absolute magnitude $M_J$ = 6.97. Thus the system 
is at a distance of 53.4 $\pm$ 4.1 pc.

\section{Chromospheric activity}
\label{halpha}

$H_\alpha$ is a chromospheric activity indicator for M dwarfs.
Stars showing a $H_\alpha$ emission line are indicated by the additional letter ``e'' in the spectral type 
column of Table~5. Although the size of our sample is too small to have a strong 
statistical sense, we find a correlation between the activity and the spectral type. The number of
active stars is greater among late-type stars than early-type stars and is increasing across the
spectral type sequence. None of the stars with spectral type earlier than M3 (over 14 objects) shows an
$H_\alpha$ emission line, whereas all of our three stars later than M8 show evidence of activity.
24\% of the M3.5-M4 stars (over 21 objects), and 47\% of the M4.5 to M6 stars (over 16 objects) show 
activity. This is consistent with the fraction of objects with $H_\alpha$ in emission versus spectral 
type reported by \cite{gizis2000} from a larger sample of M dwarfs.

\section{Conclusion}
\label{conclusion}
We obtained spectral types and spectroscopic distances for 59 high proper motion stars. These stars are 
nearby candidates, as seen from their photometric distances. 42 stars have indeed spectroscopic distances 
below the 25~pc limit of the Catalogue of Nearby Stars. 
These new neighbours are few in regards to  the large number of missing stellar systems within 25 pc ($\sim$ 2000) according to \citet{henry2002}. However, this study, joined to similar ones, allows to increase, step by step, the completeness of the solar neighbourhood sample and to cross-check
the new nearby candidates in different samples (see e.g. Table 1). Nevertheless, it is clear that further efforts are needed in order to reach a much higher completeness level.

The neighbours recovered from our study are mainly M dwarfs. Three are white 
dwarfs, cooler than 6200~K. One star, APMPM J0541-5349, appeared to be a binary consisting of two 
M2 stars with an angular separation of 3.33\arcsec, at a distance of 53.4 $\pm$ 4.1 pc from the 
Sun. We computed the radial velocity of the subdwarf candidate LHS 73 and found a space motion 
compatible with that of an object belonging to an old population. With our determined distance of 18.7 pc,
LHS 73 is among the few subdwarfs within 20 pc. Furthermore, LHS 73 has a companion, LHS 72. LHS 72/LHS 73 is 
probably the closest known pair of subdwarfs.
Two stars lie within 10~pc: APMPM J0103-3738 is a M8 star at distance d = 10.0 $\pm$ 0.8 
pc, and LP 225-57 a M4 star at d = 8.3 $\pm$ 2.8 pc. 

 The new sources that we identified are worthy 
of more detailed observations, including high-resolution imaging to search for low-mass companions, and trigonometric parallax measurements. We plan more follow-up for the nearest objects and hope to place the most promising targets on new or existing parallax programs such as that of the Cerro Tololo Inter-American Observatory Parallax Investigation \citep[CTIOPI]{jao2005}.

\section*{Acknowledgments}
C\'eline Reyl\'e acknowledges help during the observations by Olivier Hainaut and
the NTT team at the European Southern Observatory. This research has made use of the 
VizieR database, operated at CDS, Strasbourg, France.

\label{lastpage}

\end{document}